\documentstyle[12pt]{article}
\begin{document}
\title{BRST quantization of anomalous gauge theories}
\author{Nelson R.F.Braga\thanks{braga@vms1.nce.ufrj.br} \\
Instituto de F\'\i sica\\
Universidade Federal do Rio de Janeiro\\
Caixa Postal 68528\\
21945 Rio de Janeiro\\
Brasil}
\maketitle
\abstract

\noindent
It is shown how the BRST quantization can be applied to a gauge
invariant sector of theories with anomalously broken symmetries.
 This result is used to show that shifting the anomalies to a
classically  trivial sector of fields (Wess-Zumino mechanism)
makes it possible to quantize the physical sector using
a standard BRST procedure, as for a non anomalous  theory.
The trivial sector plays the role of a
topological sector if the system is quantized without shifting the
anomalies.

\vspace{2cm}

\newpage
\section{Introduction}
When considering the quantization of an anomalous gauge theory, one
possible approach is to quantize the theory without restoring gauge
invariance, as was done by Jackiw and Rajaraman\cite{JR} for the case
of the Chiral Schwinger model. A unitary, though non gauge invariant,
theory is obtained for this particular solvable model. A more general
approach is to use the Wess-Zumino mechanism, usually interpreted as
restoring  gauge invariance.
Following this approach, one can then use a
BRST quantization procedure \footnote{see for example \cite{HT} and
\cite{MA1}}. The main purpose of this article is to discuss a physical
interpretation of the Wess-Zumino(WZ) fields, that are introduced to
restore gauge invariance, by making a connection with the
quantization of the so called topological field theories. We will
make use of the Batalin Vilkovisky(BV)\cite{BV} Lagrangian BRST
quantization scheme ( also called field-antifield quantization)
because it provides, through the master equation, a systematic way of
calculating quantum contributions (anomalies and WZ terms),
but our interpretation of the WZ fields is valid for a general BRST
quantization.

The quantization of a purely quantum field theory (vanishing
classical limit) was used by Labastida et all\cite{LPW} as an
interesting approach to generate a topological 2D quantum gravity.
The starting points are just the fields and their associated
symmetries. At the classical level the Lagrangian is zero.
  The quantum action corresponds just to the gauge fixing
of the initial symmetry.  For the particular case of 2D topological
quantum gravity, enlarging the usual symmetry of 2D gravity by
including the shift symmetry
renders a non trivial
ghost structure involving a second generation of ghosts associated to
the non independence of the transformations.  At this stage, without
coupling the theory to other sectors there is no nontrivial BRST invariant
observable.

The so called Wess-Zumino mechanism is a well known technical way of
translating the anomalous breaking of  classical gauge invariance to
the appearance of new dynamical degrees of freedom, the so called
Wess-Zumino fields, originally proposed by Faddeev and
Shatashvili\cite{FS}. The new enlarged system is invariant under the
original symmetries at the quantum level and one usually says that
gauge symmetry is restored.  The Batalin-Vilkovisky (BV)
formalism\cite{BV} , also called Field-Antifield formalism provides a
powerful framework for the BRST quantization of gauge theories.  A
first discussion about the application of the BV formalism to
anomalous gauge theories was carried out by Troost et all \cite{TPN}.
In this reference it was shown how can one regularize  a theory in
order to make sense of the terms of order higher than zero in $\hbar$
in the master equation.  These terms will represent the purely
quantum part of the theory, that means, they will take  account of
the behavior of the path integral measure and that is why they only
make sense when the theory is regularized.  When anomalies are
present, there is no local solution to the master equation in the
standard space of fields and antifields.

In reference\cite{BM1} the BV quantization was applied to the chiral
\break Schwinger model. A non local WZ term was obtained and it was made
local by the introduction of an auxiliary (WZ) field. This procedure
is particular, since this non-local WZ term does not generally exists.
Recent investigations show that the application of BV to anomalous
gauge theories leads to the appearance of the Wess-Zumino terms if
the field-antifield space is extended by the inclusion of pairs
field-antifield associated to the broken gauge symmetries.  This
realization of the mechanism proposed in \cite{FS} in the BV framework
was first shown for the case of Chiral QCD2
in\cite{BM2} and then for general theories with a closed irreducible
gauge algebra in \cite{GP1}.  In these two articles it is  assumed
that the additional fields transform as elements of the original
gauge group and that there is no additional symmetry.

It was pointed out after by De Jonghe et all\cite{JST}   that when
one extends the field antifield space by adding fields that are not
present at the classical level  one should also take into account an
additional symmetry (shift symmetry) that rules out this fields at
classical level.  Thus, the gauge fixing part of the action should
include an additional term  involving the ghost associated to this
symmetry.  This fact corresponds to imposing the condition that the
gauge fixed action be the proper solution of the (zero order) master
equation.  Following this approach, they conclude that including WZ
terms one is in fact shifting the anomalies to these new symmetries.

Some important questions come in now.
If the anomaly is just shifted, can we say that BRST invariance is restored?
Can we BRST quantize a theory in which the anomalies are still
present ?
One could follow the approach that anomalies are really
canceled, as did Gomis and Paris\cite{GP2} and impose the properness
condition just on the quantum action. This corresponds to neglecting
the new symmetries. We will see however that following this approach
we loose an important physical interpretation for the origin of the
WZ fields.

The aim of this article is to show that, including these extra
symmetries, the WZ fields will be interpreted as coming from a
trivial sector that could also lead to topological field theories
depending on the quantization procedure.  We will also show that if a
general gauge theory has a broken sector of symmetries, we can use
BRST quantization for the other sector.  This fact is particularly
important if the anomaly is shifted to a non physical sector.

\section{Anomalous gauge theories}

 The BV quantization procedure is defined in an enlarged space of
fields and antifields, collectively denoted by $\Phi^a$ and
$\Phi^{*a}$ respectively.
The quantum action
has the general $\hbar$ expansion :

\begin{equation}
\label{action}
W(\Phi^a ,\Phi^{*a} ) =
S(\Phi^a ,\Phi^{*a} ) +
\sum_{p=1}^\infty \hbar^p M_p (\Phi^a ,\Phi^{*a} )
\end{equation}

It should satisfy the so called (quantum) master equation:

\begin{equation}
\label{Master}
{1\over 2}(W,W) = i\hbar\Delta W
\end{equation}

\noindent  where the antibracket is defined as:
 $(X,Y) = {\partial_rX\over
\partial\Phi^a} {\partial_lY\over\partial\Phi^{\ast a}}
- {\partial_rX\over \partial\Phi^{\ast a}}
  {\partial_lY\over \partial\Phi^a}$
and the operator Delta as:  $\Delta \equiv
{\partial_r\over\partial\Phi^a}{\partial_l\over\partial\Phi^\ast_a}\;$

\noindent Equation (\ref{Master}) implies that the vacuum functional,
defined by:
\begin{equation}
\label{Vacuum}
Z_\Psi = \int\prod D\Phi^a
exp\left( {i\over\hbar}
W(\Phi^a, \Phi^{*a} = {\partial\Psi\over \partial \Phi^a}\right)
\end{equation}

\noindent is independent of the gauge fixing fermion $\Psi$.
More details can be found in \cite{BV} or \cite{HT}.

The zero order term of the action $W$ : $  S (\Phi^a,\Phi^{\ast a})$
is usually called gauge fixed action  and is subject to the boundary
condition:

\begin{equation}
\label{boundary}
 S(\Phi^a,\Phi^{\ast a}= 0)= {\cal S} (\phi^{i} )
\end{equation}

\noindent where ${\cal S}$ is the classical limit of theory. The set of fields
$\Phi^a$ includes the classical fields $\phi^i$,  ghost fields $c^\alpha$
associated to the symmetries of ${\cal S} (\phi^{i})$
and possibly some additional fields necessary to have a standard
representation for the gauge conditions\cite{BV}.
The set of antifields $\Phi^{\ast a}$  contains the corresponding partners
of each of the fields.

We can rewrite the master equation (\ref{Master}) in powers of $\hbar$.
The two first powers are:

\begin{eqnarray}
\label{Master2}
(S,S) &=& 0\\
\label{Master3}
(M_1,S) &=& i \Delta S
\end{eqnarray}

\noindent we will consider theories for which the higher order contributions
$M_P (P\ge 2)$ to $W$ can be taken as zero, so we only need these two
first order terms in the master equation.

As mentioned before, we  need to regularize the theory in order to
make sense of the terms of order higher than zero in $\hbar$ in the
master equation, like $\Delta S$. We will not be concerned with the
details of the
regularization process in this article.  One can find them in the
literature\footnote{see for example \cite{TPN},\cite{GP2},
\cite{JO} and \cite{TP}}.  We will just present the general idea.

A regularized theory can be  build by introducing Pauli Villars(PV)
fields, and adding an extra term $S_{PV}(\chi^a,\chi^{\ast a},
\Phi^a)$ to $W$.  The PV fields $\chi^a$ have the same statistics as
the corresponding $\Phi^a$ but their path integral is defined in such
a way that the contributions from their loops has a relative minus
sign.  The regularization is obtained by a judicious choice of
$S_{PV}$ such that the contribution to  $\Delta S$ coming from both
sets of fields cancel.  The mass terms of the PV fields, necessary in order
to eliminate their propagators after the appropriate infinity mass
limit is taken, will break the zero order master equation $(S,S) =
0$. There is, as expected an arbitrariness in this regularization
process.

A theory is said to be anomalous when there is no local term $M_1$,
involving only the original fields of the theory that satisfies
the equation  (\ref{Master3}).
It can be seen in  references \cite{TPN}   and
\cite{GP2}  that anomalies correspond to a violation of
the master equation that can be put in the form:

\begin{equation}
\label{Anomaly}
{1\over 2}(W,W)- i\hbar\Delta W = c^\gamma A_\gamma
\end{equation}

\noindent  where $\gamma$ takes some values inside the domain of $\alpha$
(spatial integrations are, as usual, implicit).
The symmetries associated to the ghosts $c^\gamma$ are said to be broken
at the quantum level.

For a general gauge theory, when a particular regularization process
is chosen and we get a particular form of equation (\ref{Anomaly}), we
arrive at a quantum theory with two sectors of symmetries. The
broken ones (corresponding to $c^\gamma$ ) and the unbroken ones
(corresponding to the other ghosts).
We can incorporate into the theory the information about the symmetry
breaking, by defining a vacuum functional:

\begin{equation}
\label{ModVacuum}
\overline{Z}_{\overline\Psi} = \int\prod D\Phi^a \prod \delta(c^\gamma)
 exp\left( {i\over\hbar}
W(\Phi^a,{\partial\overline\Psi\over \partial \Phi^a}\right)
\end{equation}

\noindent where $\overline\Psi$ is a fermion independent of $c_\gamma$
It is easy to show that $\overline{Z}_{\overline\Psi}$ is independent of
$\overline\Psi$.
That means, we have a BRST invariant theory.  We will see in chapter
(3) that this procedure will enable us, by shifting the anomalies to
a trivial sector of fields, to build up a BRST invariant
generating functional where the  original symmetries of the
classical theory are realized.

\section{Wess Zumino Mechanism }

We can always associate to any standard field theory (not only
topological observables) an additional sector, that corresponds to
fields with zero Lagrangian at the classical level, in the same
spirit of refe-\break rence\cite{LPW}.
We consider a general gauge theory with extended BV action:

\begin{equation}
\label{extended action}
S = S_{Phys.}(\Phi^a,\Phi^{\ast a}) + S_T(\vartheta^b,\vartheta^{\ast b}
, c^\alpha)
\end{equation}

Subject to the boundary conditions (classical limit):

\begin{eqnarray}
\label{boundary2}
 S_{Phys.}(\Phi^a,\Phi^{\ast a}= 0)&=& {\cal S} (\phi^i)\nonumber\\
 S_{T}(\vartheta^b,\vartheta^{\ast b} = 0,c^\alpha) &=& 0
\end{eqnarray}

The set $\vartheta^b$ includes at least the fields  $\theta^\beta$
and the ghosts $d^\beta$. The classical theory is invariant
under two independent groups of gauge transformations

\begin{eqnarray}
\label{shift}
\delta \phi^i &=& R^i_\alpha (\phi^i) \lambda^\alpha         \nonumber\\
\delta \theta^\beta &=& \rho^\beta
\end{eqnarray}

\noindent where   $\lambda_\alpha$ and $\rho^\beta$ are arbitrary
functions.

We will call the first set of transformations as physical symmetries
because they are  manifest symmetries of the classical action ${\cal S}
(\phi^i )$ that we want to quantize.  We want to  build up a quantum
version of this theory that is also gauge invariant with respect to
these symmetries.  They will be fixed by the ghosts $c^\alpha $.

The invariance of the classical theory with respect to these physical
symmetries leads to Ward identities relating the Green functions and
thus the renormalization parameters, that are of extreme importance
in proving the renormalizability of the quantum theory (\cite{ZJ}).
For the case of anomalous gauge theories the Ward identities have
higher order corrections (in loops) that may spoil the
renormalizability.  One can see for example in (\cite{GJ}) and
(\cite{BIM}) that anomalies constitute an obstacle to the proof of
renormalizability for gauge theories and that this proof depends on
the ability of canceling them out, by, for example, adding extra
fermionic fields.  The addition of the Wess Zumino fields at quantum
level will also give extra contributions to these identities since
the WZ fields will also transform with the physical symmetries.
Anyway, we see that these symmetries have an important role when
considering the quantization of the physical action.

 The second set in (\ref{shift}) will be called non physical
symmetries because they just represent the absence of the fields
$\theta^\beta$ at the classical level.  The ghosts $d^\beta$ will play
the role of gauge fixing these symmetries.  When we realize the Wess
Zumino mechanism some of these symmetries will be broken, simply
reflecting the fact that, at the quantum level, the theory will no
more be independent of the WZ fields.
These symmetries are not manifest at classical level and are thus not
relevant for considering the quantization of ${\cal S}
(\phi^i )$.
That is why, as we will see at the end of this chapter, we will build
at a generating functional that does not involve this non physical
symmetries.

 There is actually not an unique way to
express the transformations for the $\theta^\beta$ fields.  The important
thing is that they eliminate their degrees of freedom at the
classical level. If we include in the second group of transformations
(\ref{shift}) additional factors associated to usual physical
symmetries (like diffeomorphism ) what happens is that we possibly get non
independent gauge transformations, leading to the introduction of
higher order ghosts, like in \cite{LPW} .  The presence of $c^\alpha$
(ghosts associated to the symmetries of ${\cal S}(\phi^i)$- that we
will call physical symmetries) in $S_T$ is associated to the
arbitrariness in the transformation of $\theta^\beta$ with respect to
this  gauge group, since these fields are  not present at the
classical level.

We can assume that the trivial sector contains a set of fields
that have  the same structure (Lorentz plus internal symmetries)  of
the elements of the physical gauge group.
Now, introducing the Pauli-Villars fields to regularize the physical
sector, we may choose different
mass terms that may break some original physical symmetries, some
symmetries of the trivial sector or, in general, a linear combination
of them\cite{JST}.
We prefer to consider a choice of mass terms that do not involve
the fields $\theta^\beta$ and thus will break only symmetries of
the physical sector . Assuming that the new fields in $S_T$ have an
invariant path integral measure, we  get

\begin{equation}
\label{DeltaS}
\Delta S = \Delta S_{Phys.} = c^\gamma A_\gamma
\end{equation}

Following the idea of \cite{BM2} and \cite{GP1} we can write out a
quantum contribution $M_1(\phi^i, \theta^\beta)$ that cancels the
contribution of (\ref{DeltaS}) to the master equation(\ref{Master}).
We know that this $M_1$ must depend on  the extra fields
$\theta^\beta$ because we are assuming that the theory has genuine
anomalies and, as is well known, they can not be canceled by just
counter terms.  Therefore $M_1$ is not invariant under (\ref{shift}),
leading to a violation in the master equation that now has the
general form:

\begin{equation}
\label{Anomaly2}
{1\over 2}(W,W)- i\hbar\Delta W = d^\gamma \overline A_\gamma
\end{equation}

\noindent we say now that we have implemented the Wess-Zumino mechanism.

The anomalies have not been canceled. They have just been shifted to
the symmetries associated to the trivial sector. We can define again
in the same spirit of (\ref{ModVacuum})

\begin{equation}
\label{Vacuum2}
\overline{Z}_\Psi = \int\prod D\Phi^a \prod D\vartheta^b
\prod \delta(d^\gamma)
exp\left( {i\over\hbar}
W(\Phi^a,{\partial\overline\Psi\over \partial \Phi^a}, \vartheta^b,
{\partial\overline\Psi\over \partial \vartheta^b}
\right)
\end{equation}

\noindent where $\overline\Psi$ is a fermion independent of $d^\gamma$.
Now the functional $\overline{Z}_{\overline\Psi} $ involves integrations over
the whole set of physical fields (all the ghosts $c^\alpha$ are included).
We can couple the fields to sources $J$ and also introduce the
sources $L$ writing a generating functional:

\begin{eqnarray}
\label{generator}
\overline{Z}[ J^a,J^b,L^a,L^b ]_\Psi  = \int \prod  D\Phi^a
\prod D\vartheta^b \prod \delta(d^\gamma)
\nonumber \\  exp\left( {i\over\hbar}
W(\Phi^a,{\partial\overline\Psi\over \partial \Phi^a} +L^a, \vartheta^b,
{\partial\overline\Psi\over \partial \vartheta^b} +L^b )
+ J^a \Phi^a + J^b \Phi^b
\right)
\end{eqnarray}
\noindent defining the classical fields and effective action
respectively as:

\begin{eqnarray}
\label{effective}
\phi^A_{cl} &=& {\hbar\over i}{\delta ln \overline{Z}[J^A,L^A]_\Psi
\over \delta J^A}\\
\label{Action}
\Gamma_\Psi [\phi^A, L^A] &=& {\hbar\over i}ln \overline{Z}_{\Psi} -
J^A\phi^A_{cl}
\end{eqnarray}
\noindent with A=(a,b).

The Zinn-Justin equation

\begin{equation}
\label{ZJ}
(\Gamma_{\overline\Psi} , \Gamma_{\overline\Psi} ) = 0
\end{equation}

\noindent (with the antibracket defined in a space where $\phi^A$
play the role of the fields and $ L^A$ of the antifields)
will now express the gauge invariance of the physical
sector and possibly some trivial uncoupled symmetries of the trivial
sector.

If we do not include  $M_1(\phi^i, \theta^\beta)$ in the quantum action
the anomaly will show up in the physical sector. The trivial sector
will remain uncoupled and may lead to topological theories like
in\cite{LPW}.  At least for some simple models, like in \cite{JR},
one can then possibly quantize the physical sector in a non gauge
invariant way and proceed the BRST quantization for the trivial
sector, like in\cite{LPW}.

Gauge invariance is of extreme importance in proving the
unitarity\cite{GT} of  field theories .
Implementing the  Wess Zumino mechanism as in the present  chapter we
get  a non anomalous version for the potentially anomalous theory
\footnote{ Helpful discussions about anomalies and their physical
implications can be found for example in \cite{J} and \cite{P}}.
In the BRST language, we get a quantum theory with a nillpotent
BRST generator, representing the invariance of the effective action
(\ref{Action}).  One can then define the physical states in the usual
way \cite{HT}, in terms of the cohomology classes of this generator.
If gauge invariance is lost the theory may become non  Unitary, and
therefore inconsistent.

One could also expect, in principle, that the renormalization
properties are improved by restoring gauge invariance.  We
can see, however, in \cite{DF} an example of a four dimensional gauge theory
that after the decoupling of one of the  fermion chiralities remains
gauge invariant by the generation of a Wess Zumino term, but is
non renormalizable. So, the issue of renormalizability can not
be analised in a general way by just taking gauge invariance into account.

\section{Example}

Now we will consider an  example in order to illustrate
our previous development. Let us consider a theory
that at the classical level is described by the sum of the following actions:

\begin{eqnarray}
S_{Phys.} &=& \int d^2x \{i \overline\psi\; {\slash\!\!\!\!D} \;
{(1-\gamma_5)\over 2}\;\psi
- {1\over 4} F_{\mu\nu}F^{\mu\nu}
 + A^\ast_\mu \partial^\mu c \nonumber\\
\label{SP}
&+& i\psi^\ast\psi c - i\overline\psi \;\overline{\psi^\ast} c \}
\\ \label{ST}
S_T &=& \int d^2x \{ \theta^* c + \theta^* d + \overline c^* \pi +
\overline d^* \lambda \}
\end{eqnarray}

The action $S_{Phys.}$ corresponds to the gauge fixed BV action for the
chiral Schwinger model\cite{BM1} and $S_T$ corresponds to the gauge
fixed action for a theory of a scalar field that transforms with
the gauge group of the Schwinger model (corresponding to the
ghost c) and also with an additional symmetry (corresponding to the
ghost d). The antighosts $\overline c$ and $\overline d$ are introduced
in order to allow the implementation of the gauge choices in the standard
BV way : $\Phi^{*a} = {\partial\Psi\over \partial \Phi^a}$.

The boundary conditions satisfied by  $S_{Phys.}$ and $S_T$ are of the
same form as in (\ref{boundary2}) with ${\cal S} (\phi^i)$ being, in
this case, the classical action for the chiral Schwinger model.

The BRST transformations for the fields in  $S_T$ are:

\begin{eqnarray}
\delta c &=& 0
\,\,\,\;\;\; ,\,\,\,\,\delta d = 0\nonumber\\
\delta \theta &=& c+d\;\;\;\,\,\, ,
\delta \overline c = \pi\nonumber\\
\delta \overline d &=& \lambda\;\;\;\,\,\, ,
\,\,\delta \pi = 0\nonumber\\
\delta \lambda &=& 0
\end{eqnarray}

 We can write $S_T$  as a BRST variation, showing explicitly it's
topological character:

\begin{equation}
S_T = \delta \Omega
\end{equation}

\noindent with

\begin{equation}
\Omega = - \theta^* \theta + \overline c^* \overline c +
 \overline d^* \overline d
\end{equation}

To implement the BV quantization for $S = S_{Phys.} + S_T $
this theory must be regularized before the calculation of $\Delta S$.
Let us consider first  $S_T$. In
order to regularize this action we can consider , as already discussed in
section (2), the  Pauli Villars (PV) regularization.  We
have to include a Pauli Villars partner for the field $\theta$, with the
same kinetic operator but with a mass term in such a way that after
taking the infinity mass limit would have a vanishing propagator.

The problem is that, contrarily to reference (\cite{TPN}), our $\theta$ field,
has no kinetic term at classical level. Thus, $\theta$ itself has a
vanishing propagator at the classical level.
We can overcome this difficulty, for example, by considering an action

\begin{equation}
S_T(\alpha)  =  S_T + \int d^2x \{ \alpha\partial_\mu\theta
\partial^\mu \theta \}
\end{equation}

\noindent The extra kinetic term breaks the gauge invariance, as can
be seen from:

\begin{equation}
\label{Reg}
\big( S_T(\alpha), S_T(\alpha)\big)  = -2\alpha\Box \theta ( c + d )
\end{equation}

\noindent but in the limit $\alpha\rightarrow 0$ we recover the
original theory with it's invariances.
We can now introduce a PV
partner to $\theta$, say $\overline \theta$, with the same vanishing classical
limit , in the same spirit of (\cite{TPN}):

\begin{equation}
S_{PV}(\alpha)  = \int d^2x \{ \alpha\partial_\mu\overline\theta
\partial^\mu \overline\theta + M^2\overline\theta^2 +
\overline\theta^* (c+d) \}
\end{equation}

\noindent The violation of the master equation now takes the form:

\begin{equation}
\big( S_T(\alpha) + S_{PV}(\alpha), S_T(\alpha) + S_{PV}(\alpha) \big)
= -2\alpha\Box (\theta +  \overline\theta ) ( c + d ) + M^2 \phi ( c + d )
\end{equation}

Functionally integrating  the PV field $ \overline\theta$ it's
contribution to the above expression just vanish, because, contrarily
to the cases considered in  (\cite{TPN}), the field $ \overline\theta$
is present only in the kinetic term. So, we recover (\ref{Reg}).
Taking then the limit of vanishing $\alpha$ we find that the
contribution of $S_T$ to $\Delta S$ is zero.  This was clearly
expected from the fact that the generators of the symmetries of the
field $\theta$ are field independent.

On the other hand, the action  $ S_{Phys.}$ is exactly the same action
that was considered in \cite{BM1}, where the following result was
obtained:

\begin{equation}
\label{Delta2}
\Delta S_{phys.} = {i\over 4\pi}\int d^2x\,\; c\; [ (1-a)\partial_\mu A^\mu -
\epsilon^{\mu\nu}\partial_\mu A_\nu ]
\end{equation}

Now  we must build up a quantum action $W$ of the form
of eq. (\ref{action}), whose first component is just  $S = S_{Phys.} + S_T $.
We can consider  two different approaches . The first one is to take all
the higher order contributions $M_P$ to the action (\ref{action}) as
vanishing. Then eq.
(\ref{Delta2}) implies a violation of the master equation of the same
form as (\ref{Anomaly}). In this case, following the lines of
section (2), we may just take the symmetry associated to the ghost $c$
out of the BRST setting  by considering the vacuum functional
$\overline{Z}_{\overline\Psi}$ of (\ref{ModVacuum}).  The two sectors,
associated to $S_{Phys.}$ and  $S_T$ will then remain uncoupled. The
quantization of  $S_{Phys.}$  can be performed exactly as in
\cite{JR}, where the chiral Schwinger model was shown to contain a free
massive vector boson plus harmonic excitations.
 while the  sector corresponding to $S_T$ will now
correspond to the action (removing also the antighosts and auxiliary
fields associated to $c$ )

\begin{equation}
\overline S_T (\phi^a,\phi^{\ast a})  = \int d^2x \{ \theta^* d +
\overline d^* \lambda  \}
\end{equation}

\noindent that represents a scalar field
with no non trivial BRST
invariant observable, as it happens in \cite{LPW} for topological 2D
gravity. This sector can possibly  be coupled to other topological theories
in order to generate non trivial observables. We can gauge fix this
action  by choosing, for example, the scalar field to
be equal to some preferable field $\theta_o$, introducing the fermion:

\begin{equation}
\Psi = \overline d (\theta - \theta_o )
\end{equation}

\noindent that leads to the action

\begin{equation}
\overline S_T (\phi^a,\phi^{\ast a}= {\partial\Psi\over \partial\phi^a})
  = \int d^2x \{ \overline d d +
 \lambda (\theta - \theta_o )\}
\end{equation}

A different approach to quantize the theory described by  (\ref{SP}) plus
(\ref{ST})
is to implement the Wess Zumino mechanism,
following the lines of section (3). In ref. \cite{BM1} it was shown that
adding the contribution

\begin{equation}
\label{M1}
 M_1 = -{1\over 4\pi} \int d^2x \left\lbrace {(a-1)\over 2}\,
\partial_\mu \theta \; \partial^\mu \theta + \theta \left\lbrack (a-1)
\partial_\mu A^\mu + \epsilon^{\mu\nu} \partial_\mu A_\nu \right\rbrack
\right\rbrace
\end{equation}

\noindent to the quantum action of the Schwinger model
would cancel the contribution from $\Delta S$ to the master equation
if the term $\theta^* d$ is not present in the classical action $S$.
The inclusion of this extra term, taking into account, as already
explained, the additional symmetry associated to the ghost d, leads
to the following result:

\begin{equation}
(M_1,S) = i \Delta S_{Phys.} +\int d^2x \overline A(\theta,A_\mu) d
\end{equation}

\noindent with

\begin{equation}
\overline A(\theta,A_\mu) = {1\over 4\pi}\Big( (a-1) \Box \theta  -
\partial_\mu ((a-1)A_\mu +\epsilon^{\mu\nu}A_\nu)  \Big)
\end{equation}

\noindent Now the quantum action $W=  S_{Phys.} + S_T + \hbar M_1$
satisfies  the equation
\begin{equation}
{1\over 2}(W,W)- i\hbar\Delta W = d^\gamma \overline A_\gamma
\end{equation}

\noindent representing the fact that the introduction of the Wess-Zumino
term $M_1$
has shifted the anomaly from the physical symmetry, associated with $c$
to the non physical symmetry associated to the ghost $d$ (only
present in $S_T$ ).
Following chapter (3) we then define a
vacuum functional as in (\ref{Vacuum2}) that takes into account the
breaking of the symmetry associated to the ghost d, ruling out this
field from the formulation. The theory will then be described by the action

\begin{equation}
\overline W = S_{Phys.} + \int d^2x \{ \theta^* c +  \overline c^* \pi \}
+ M_1 (A_\mu , \theta )
\end{equation}

\noindent that corresponds to the
Schwinger model with it's standard Wess Zumino term.
That means, we arrive at a gauge invariant formulation (with respect to
the physical symmetry ) for the theory.

Thus we see from this simple example that the Wess Zumino term
corresponds to the coupling of the physical sector to a trivial sector
that otherwise would play the role of a topological sector.

\section{Conclusions}
It is interesting now to make a parallel with the original discussion
of ref\cite{FS}.  There, the anomalies are interpreted as not
breaking the gauge symmetry but just inducing a different
representation for the group, in which the WZ fields are also
present.  We can say that in order to build up this representation
for the gauge group we are borrowing some fields from a sector that
was in principle trivial. In fact the name trivial is not so
appropriated.  We have seen that, although trivial at classical
level, depending on the quantization process this sector may lead to
Topological Field Theories at quantum level if we don't use it to
implement the Wess-Zumino mechanism.  We have learned from recent
studies\cite{W}, \cite{BB} that some interesting results can emerge
from this kind of theories.

Regarding the WZ mechanism as a breaking in the symmetry of what
would be a topological sector leads us to some interesting questions
for future investigations.  If part of the symmetry of this sector is
broken in the quantization process by coupling to another sector we
get possibly a mechanism for generating field theories with not only
topological observables beginning with topological invariant actions.
In other words we get a mechanism for coupling topological theories
to  non topological ones.

It is worth mentioning that it has been considered by Marnelius the
possibility of  BRST quantization of anomalous field
theories\cite{MA2}.  This author considers BRST quantization with
$Q_{BRST}^2 \neq 0$ but with $Q_{BRST}$ conserved.  What we were
interested in discussing was exactly the mechanism of ``restoring
gauge invariance'', so what we did was to exclude the broken
symmetries from the BRST setting. Thus we can define a nillpotent and
conserved BRST charge. In the case of (\ref{Vacuum2}) this
corresponds to the charge that generates the BRST transformations on
the physical sector plus some possible remaining (uncoupled)
symmetries of the trivial sector.  After the master equation is
written in the form of (\ref{Anomaly2}) we can compute the generator
of the BRST transformations excluding the ones associated to the
ghosts $d^\gamma$ ( that means the BRST symmetries of
(\ref{Vacuum2})) : $\overline{Q}_{BRST}$.  The physical states will
be defined by:
\begin{equation}
\label{states}
\overline{Q}_{BRST} \vert Phys. > = 0
\end{equation}
\noindent All the standard BRST procedure\cite{HT} \cite{MA1} can then
be applied.

Although the present analysis is based on the BV quantization framework,
we can generalize our interpretation about the origin of the
Wess-Zumino fields for general BRST quantization.

\noindent ACKNOWLEDGEMENTS: The author would like to thank
\break R.Amorim, J.Barcelos-Neto, H.Boschi Filho, A.Das, E.C.Marino,\break
J.A.Mignaco and
C.Wotzasek for important discussions and \break F.De Jonghe for important
 correspondence. The author is partially supported by CNPq-Brazil.


\begin{thebibliography}{30}
\bibitem{JR} R. Jackiw and R. Rajaraman, Phys. Rev. Lett. 54 (1985) 1219.
\bibitem{HT}M.Henneaux and C.Teitelboim, Quantization of Gauge Systems,
Princeton University Press 1992, Princeton, New Jersey.
\bibitem{MA1}R.Marnelius, Nucl.Phys.B395 (1993) 647, B391 (1993) 621,
B384 (1992) 318; S.Hwang and R.Marnelius, Nucl.Phys.B315 (1989) 638.
\bibitem{BV} I. A. Batalin and G. A. Vilkovisky, Phys. Lett. B102
(1981) 27, Phys. Rev. D28 (1983) 2567.
\bibitem{LPW}J.M.F.Labastida, M.Pernici and E.Witten;
 Nucl.Phys.B310, 611 (1988).
\bibitem{FS} L.D.Faddeev, Phys. Lett. B145 (1984) 81;
 L.D.Faddeev and S.L.Shatashvili, Phys. Lett. B167 (1986) 225.
\bibitem{TPN}  W.Troost, P.van Nieuwenhuizen and
A. Van Proeyen, Nucl. Phys. B333 (1990) 727.
\bibitem{BM1} N.R.F.Braga and H.Montani, Phys. Lett. B264 (1991) 125.
\bibitem{BM2} N.R.F.Braga and H.Montani, Int. J. Mod. Phys. A8 (1993)
2569.
\bibitem{GP1} J. Gomis and J. Paris, Nucl. Phys. B 395 (1993) 288.
\bibitem{JST} F.De Jonghe, R.Siebelink, W.Troost, Phys.Lett.B306(1993)295.
\bibitem{GP2} J.Gomis and J.Paris, Anomalies and Wess-Zumino terms in
an extended, regularized Field-Antifield formalism, Preprint
UB-ECM-PF 93/14, HEP-TH 9401161.
\bibitem{JO}F.De Jonghe, The Batalin-Vilkovisky Lagrangian Quantization
scheme with applications to the study of anomalies in gauge theories,
PH.D. thesis K.U. Leuven, HEP-TH 9403143.
\bibitem{TP} Regularization and the BV formalism, W. Troost and A.Van
Proeyen, presented at "Strings 93", Berkeley, May 1993.
\bibitem{ZJ}Zinn-Justin,J. "Renormalization of gauge theories", in
Trends in Elementary Particle Theory, Lecture Notes in Physics Vol. 37,
ed. H.Rollnik and K.Dietz, Berlin, Springer.
\bibitem{GJ}D.J.Gross and R.Jackiw , Phys.Rev.D6 (1972) 477.
\bibitem{BIM}C.Bouchiat, J.Iliopoulos and Ph. Meyer, Phys.Lett.B38
(1972) 519.
\bibitem{GT} G.'t Hooft, Nucl.Phys. B33 (1971) 173, B35 (1971) 167.
\bibitem{J}R.Jackiw , ``Anomalies and Topology", Proceedings of Theoretical
Advanced Study Institute in Elementary Particle Physics, Yale ASI 1985:83
(QCD161 : Y2 : 1985 ).
\bibitem{P} J. Preskill, Ann. Phys. 210 (1991) 323.
\bibitem{DF}E.D'Hoker and E. Farhi, ``Effective Actions, Decoupling and
Anomalies in Chiral Gauge Theories", Symposium on Anomalies Geometry
Topology, Eds. W.A.Bardeen and A.White, World Scientific 1985.
\bibitem{W}E.Witten, Comm. Math. Phys. 117(1988)353.
\bibitem{BB} For a general review on the subject see: D.Birmingham,
M.Blau, M.Rakowski and G.Thompson, Phys.Rep.209(1991) 129.
\bibitem{MA2}R.Marnelius, Nucl.Phys.B315(1989)638.
\end{thebibliography}
\end{document}